\documentclass[fp,twocolumn]{jpsj3}
\usepackage{txfonts}
\usepackage{here}
\usepackage{graphicx}
\usepackage{graphics}
\usepackage{color}

\title{Pressure-Induced Metallization in Iron-Based Ladder Compounds Ba$_{1-x}$Cs$_x$Fe$_2$Se$_3$}

\author{Takafumi Hawai$^{1,2}$, Chizuru Kawashima$^3$,  Kenya Ohgushi$^4$, Kazuyuki Matsubayashi$^5$, Yusuke Nambu$^6$, Yoshiya Uwatoko$^2$, Taku J. Sato$^7$, and Hiroki Takahashi$^3$}

\inst{$^1$ High Energy Accelerator Research Organization (KEK), Tsukuba, Ibaraki 305-0801, Japan \\
$^2$Institute for Solid State Physics, University of Tokyo, Kashiwa, Chiba 277-8581, Japan \\
$^3$ College of Humanities \& Sciences, Nihon University, Sakurajosui, Setagaya-ku, Tokyo 156-8550, Japan \\
$^4$Department of Physics, Tohoku University, Sendai, Miyagi 980-8578, Japan \\
$^5$Department of Engineering Science, The University of Electro-Communications, Chofu, Tokyo 182-8585, Japan \\
$^6$Institute for Materials Research, Tohoku University, Sendai, Miyagi 980-8577, Japan \\
$^7$Institute of Multidisciplinary Research for Advanced Materials, Tohoku University, Sendai, Miyagi 980-8577, Japan}

\abst{
Electrical resistivity measurements have been performed on the iron-based ladder compounds Ba$_{1-x}$Cs$_x$Fe$_2$Se$_3$ ($x$ = 0, 0.25, 0.65, and 1) under high pressure.
A cubic anvil press was used up to 8.0 GPa, whereas further higher pressure was applied using a diamond anvil cell up to 30.0 GPa. 
Metallic behavior of the electrical conductivity was confirmed in the $x$ = 0.25 and 0.65 samples for pressures greater than 11.3 and 14.4 GPa, respectively, with the low-temperature $\log T$ upturn being consistent with weak localization of 2D electrons due to random potential. 
At pressures higher than 23.8 GPa, three-dimensional Fermi-liquid-like behavior was observed in the latter sample.
No metallic conductivity was observed in the parent compounds BaFe$_2$Se$_3$ ($x $ = 0) up to 30.0 GPa and CsFe$_2$Se$_3$ ($x$ = 1) up to 17.0 GPa.
The present results indicate that the origins of the insulating ground states in the parent and intermediate compounds  are intrinsically different; the former is a Mott insulator, whereas the latter is an Anderson insulator owing to the random substitution of Cs for Ba.
}

\begin{document}
\maketitle

\section{Introduction}

\begin{figure}[b]
\centering
\includegraphics[width=0.8\linewidth]{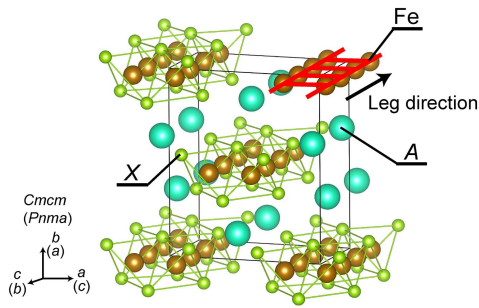}
\caption{(Color online) Crystal structure of iron-based ladder compounds $A$Fe$_2 X_3$. Red lines emphasize the ladder structure of iron atoms. The arrow indicates the leg direction.
}
\label{crystal_structure}
\end{figure}

\begin{figure*}[!h!t!b]
\includegraphics[width=0.9\linewidth]{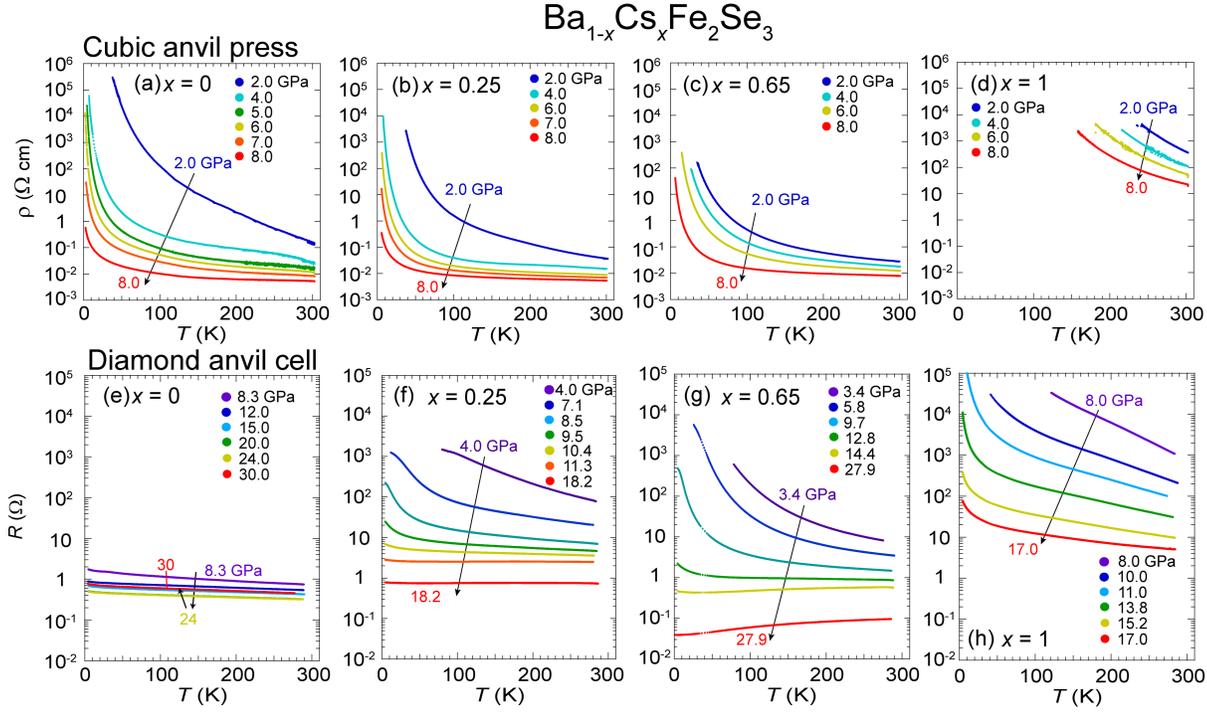}
\caption{(Color online)  Temperature dependence of the electrical resistivity $\rho$ and electrical resistivity $R$ along the leg direction at high pressure  for Ba$_{1-x}$Cs$_x$Fe$_2$Se$_3$ ($x = $0, 0.25, 0.65, and 1): (a)-(d) high pressure is applied using a cubic anvil press; (e)-(h) high pressure is applied using a diamond anvil cell.}
\label{Results_of_all}
\end{figure*}

Iron-based ladder compounds $A$Fe$_2X_3$ ($A$ = K, Rb, Cs, and Ba; $X$ = S, Se, and Te)  \cite{Hong197293, Klepp19961, Mitchell20041867} attract considerable attention as low-dimensional analogues of iron-based superconductors.
Figure \ref{crystal_structure} shows crystal structure of iron-based ladder compounds \cite{Momma:db5098}.
The compounds have quasi-one-dimensional two-leg Fe ladders separated by the $A$ cations.
The two-leg ladder is formed by edge-sharing [Fe$X_4$] tetrahedral structures, of which the connectivity is the same as those found for the two-dimensional square lattice of Fe in iron-based superconductors.
Earlier neutron diffraction experiments show that at low temperatures, several different magnetic structures are stabilized depending on $A$ and $X$.
For instance, block magnetism, in which four magnetic moments on Fe atoms form a ferromagnetic block and each block is aligned antiferromagnetically along the leg direction, was observed in BaFe$_2$Se$_3$ \cite{PhysRevB.84.214511, PhysRevB.84.245132, PhysRevB.84.180409, PhysRevB.85.064413} and stripe magnetism was observed in $A$Fe$_2$$X_3$ ($A$ = K, Cs, Ba, $X$ = S, Se) \cite{ takahashi2015pressure, PhysRevB.85.214436, PhysRevB.85.180405}.
These magnetic structures are also analogues to the ones found in the iron-based superconductors, such as the block magnetic structure in the 245 system and the single stripe structure in 1111 systems.

Despite the similarity in the [Fe$X_4$] edge-sharing connectivity and also in the magnetic structures, the compounds are insulators, in contrast to the metallic nature of iron-based superconductors.
Therefore, the metallization of iron-based ladder compounds has been highly desired.
Carrier doping is one plausible way of realizing metallization.
However, previous studies on hole-doped \cite{PhysRevB.85.180405, 0953-8984-26-2-026002, PhysRevB.91.184416, PhysRevB.92.205109} and electron-doped \cite{PhysRevB.90.085143 ,PhysRevB.92.205109} iron-based ladder compounds concluded that the compounds are insulating for all compositions.
Several intriguing findings were, nevertheless, reported.
The study of Ba$_{1-x}$Cs$_x$Fe$_2$Se$_3$ revealed a large decrease in resistivity at intermediate compositions \cite{PhysRevB.91.184416}.
The block magnetic order is suppressed completely at $x$ = 0.25, where no magnetic signal was observed in powder neutron diffraction profiles down to 7 K.
On the other hand, the low-temperature divergence of the resistivity is most suppressed at $x$ = 0.65, indicating that the system is the closest to the metallic state among the intermediate compounds.
Another way of realizing metallization may be to apply pressure.
Indeed, Takahashi et al. reported metallization and the appearance of  superconductivity in BaFe$_2$S$_3$ under a pressure of 11 GPa \cite{takahashi2015pressure}.

Combining the two observations, i.e. the decrease in the resistivity of career-doped Ba$_{1-x}$Cs$_x$Fe$_2$Se$_3$ and the pressure-induced metallization of BaFe$_2$S$_3$, we can expect the metallization of Ba$_{1-x}$Cs$_x$Fe$_2$Se$_3$ under high pressure.
In the present study, we have performed resistivity measurements under high pressure for the parent compounds BaFe$_2$Se$_3$ and CsFe$_2$Se$_3$, and for the intermediate compounds with $x$ = 0.25 (Ba$_{0.75}$Cs$_{0.25}$Fe$_2$Se$_3$) and $x $ = 0.65 (Ba$_{0.35}$Cs$_{0.65}$Fe$_2$Se$_3$).
No superconductivity is observed in all the samples, nevertheless, our results show metallic behavior of the $x$ = 0.25 and $x$ = 0.65 samples.
It may be noted that in this work, we use ''metallic'' when the sample shows a negative temperature coefficient of resistance.
We discuss the mechanism of the insulating behavior and metal-insulator transition in iron-based ladder compounds, Ba$_{1-x}$Cs$_x$Fe$_2$Se$_3$, based on the different responses to pressure of the parent and intermediate compounds.

\section{Experimental Procedure}

Single crystals with $x$ = 0, 0.25, 0.65, and 1 were synthesized by the slow-cooling method\cite{PhysRevB.85.064413, PhysRevB.85.214436}.
In this report, $x$ denotes the nominal composition.
The samples were characterized using powder X-ray diffraction with Cu $K\alpha$ radiation (not shown), and the composition dependence of the lattice parameters almost obeys Vegard's law.
Electrical resistivity was measured by the four-probe dc technique with current flow along the leg direction in the $T$ range between 4.2 and 300 K and in the pressure range between 2.0 and 30.0 GPa.
Resistivity at ambient pressure was measured by Physical Property Measurement System  (PPMS, Quantum Design) \cite{PhysRevB.91.184416}.
A cubic anvil press consisting of tungsten carbide anvils was used to measure resistivity at pressures of 2-8 GPa.
The pressure-transmitting medium for the cubic anvil press experiments was a mixture of Fluorinert FC70 and FC77 (3M Company) with a 1:1 ratio.
Four gold wires were attached to the sample using gold paste (Tokuriki 8560).
A diamond anvil cell (DAC) was used to measure resistance at pressures of 3.8-30.0 GPa. 
The pressure-transmitting medium was NaCl powder.
Thin platinum ribbons were pressed and attached to samples as electrodes.
A rhenium gasket was used, and a thin BN layer was installed as electric insulation between the platinum electrodes and the gasket.
The cubic anvil press and DAC were placed in a cryostat and cooled with liquid $^{4}$He.

\section{Results}

Figures \ref{Results_of_all}(a)-\ref{Results_of_all}(d) show the temperature and pressure dependences of electric resistivity for $x = 0, 0.25, 0.65$, and 1, respectively, measured using a cubic anvil press.
Overall, the resistivity decreases with increasing pressure.
Upon careful inspection, we found that the parent compounds with $x$ = 0 and 1 show stronger suppression of resistivity than the mixed compounds.
The resistivity of the $x = 0, 0.25,$ and 0.65 samples reaches less than $1 \times 10^{-2} $ $\Omega$ cm at 8.0 GPa, but there is no metallization or superconductivity.
Interestingly, at 8.0 GPa, the $x$ = 0, 0.25, and 0.65 samples show similar resistivities at room temperature, although their crystal and magnetic structures at ambient pressure are different.
On the other hand, the $x$ = 1 sample still shows much higher resistivity than the other compounds at room temperature.
A structural transition was reported in a previous study \cite{0953-8984-25-31-315403} for $x$ = 0 at 6 GPa.
However, there is no corresponding anomaly in our results.

For further higher pressures, we performed resistance measurements using a DAC.
Figures \ref{Results_of_all}(e)-\ref{Results_of_all}(h) show the results of electrical resistance for $x$ = 0, 0.25, 0.65, and 1 samples under higher pressures.
As seen in Figs. \ref{Results_of_all}(e) and \ref{Results_of_all}(h), the resistance of the parent compounds with $x$ = 0 and 1 increases with decreasing temperature, indicating that the parent compounds are still insulators under pressures of 30.0 GPa and 17.0 GPa, respectively.

\begin{figure}[htb]
\centering
\includegraphics[width=0.95\linewidth]{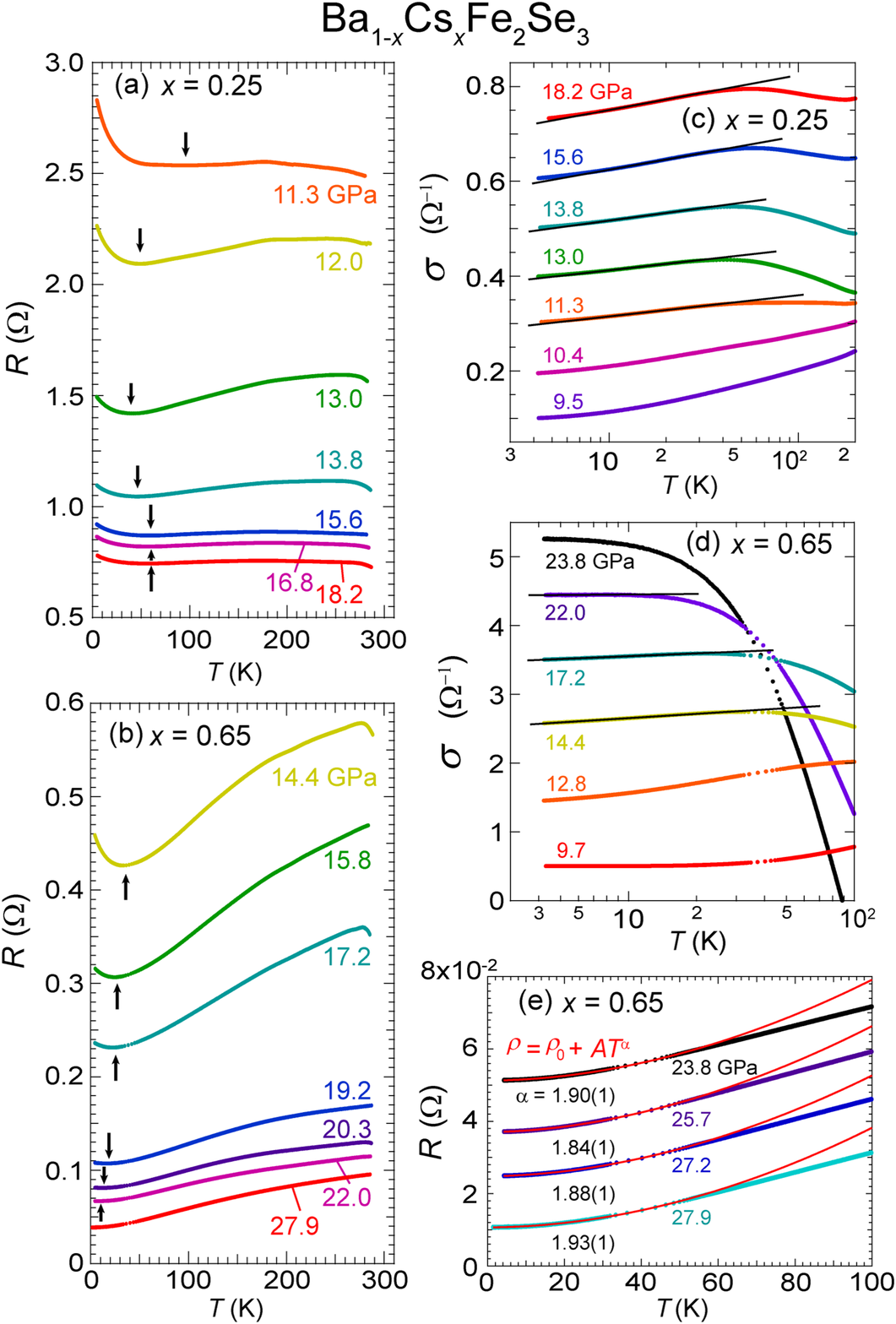}
\caption{(Color online)  (a), (b) Metallic behavior of resistance for (a) $x = $ 0.25 and (b)  $x = $ 0.65. Note that the vertical axis is a linear scale. Arrows indicate the metal-insulator transition. (c), (d) Temperature dependence on semilogarithmic scale for (c) $x = $ 0.25 at 9.5-18.2 GPa and (d) $x = $ 0.65 at 9.7-23.8 GPa. (e) Metallic behavior in the $x = $ 0.65 sample at low temperatures at 23.8-27.9 GPa. Note that all points in (c)-(e) are shifted for clarity. Solid lines show the results of fitting with (c), (d) $\sigma = \sigma_0 + A \log{T}$ and (e) the power law $\rho = \rho_0 + AT^\alpha$.
}
\label{DAC_results_allin}
\end{figure}

In contrast to the insulating behavior in the parent compounds, metallic behavior was observed in the $x$ = 0.25 and 0.65 samples, as detailed below.
Figure \ref{Results_of_all}(f) shows the pressure dependence of the resistance in the $x$ = 0.25 sample.
The resistance is greatly suppressed as the pressure is increased, and indeed, the $R$-$T$ curves become almost flat on a log scale for pressures greater than 11 GPa.
Figure \ref{DAC_results_allin}(a) shows the same $R$-$T$ curves of the $x$ = 0.25 sample on a linear scale.
As clearly seen in this figure, the resistance between 100 and 200 K decreases with decreasing temperature for pressures greater than 11 GPa, indicating that the metallic state is realized.

On further reducing the temperature, resistivity shows an upturn for all the pressures, indicating that the lowest-temperature state is again insulating for $x$ = 0.25.
Such an upturn can also be seen in the cuprate ladder compound Sr$_{2.5}$Ca$_{11.5}$Cu$_{24}$O$_{41}$ \cite{PhysRevLett.81.1090} and the two-dimensional organic conductor (DOET)$_2$BF$_4$ \cite{doi:10.1143/JPSJ.75.024701}, indicating the existence of a metal-insulator transition at lower temperatures.
The arrows in the figure denote the metal-insulator transition temperature, which was determined using d$R$/d$T$ = 0.
Applying higher pressure generally tends to suppress the metal-insulator transition; however, the insulating state for $x = 0.25$ at the lowest temperature is even robust at the highest pressure of 18.2 GPa applied in this work.
The temperature dependence of the conductivity for $x$ = 0.25 is shown in a semilogarithmic plot [Fig. \ref{DAC_results_allin}(c)].
One can see that the conductivity is proportional to $\log T$ below 50 K in the pressure range between 11.3 and 18.2 GPa, where a low-temperature insulating phase can be seen, and slightly deviates from the relation below 7 K.
In a two-dimensional system with perturbative random potential, namely, a weakly localized system, it is known that conductivity shows the relation $\sigma =  \sigma_0 + A\log T$, $A > 0$ \cite{PhysRevLett.42.673,gor1996particle, PhysRevLett.44.1288}.
Hence, the $\log T$ dependence of the conductivity suggests that this iron-based ladder compound with $x$ = 0.25 can be regarded as a disordered two-dimensional system at these pressures.

For $x =$ 0.65, metallization was achieved at 14.4 GPa [Fig. \ref{DAC_results_allin}(b)]. 
At this pressure, a metal-insulator transition similar to that of the $x$ = 0.25 sample was observed at $\sim$30 K.
Figure \ref{DAC_results_allin}(d) shows the $\log T$ dependence of the conductivity for $x$ = 0.65.
In the low-temperature insulating phase found for 14.4 $\leq$ pressure ($P$) $\leq$ 22.0 GPa, the conductivity obeys $\sigma =  \sigma_0 + A\log T$, indicating a two-dimensional feature similar to that of the $x$ = 0.25 sample.
The metal-insulator transition is completely suppressed at 23.8 GPa, where we found the system is metallic down to the base temperature.
The low-temperature part of the resistance in the range of 23.8 to 27.9 GPa is shown in Fig. \ref{DAC_results_allin}(e).
As seen in the figure, the resistance of the $x$ = 0.65 sample shows fully metallic behavior in the measured temperature range and is well expressed by the power law $\rho = \rho_0 + AT^\alpha$.
The results of the power-law fitting are also shown in Fig. \ref{DAC_results_allin}(e).
The resistivity follows the power law below 50 K.
The $\alpha$ values are almost 2, indicating that the system is in a nearly three-dimensional Fermi liquid state.

\section{Discussion}

As described above, BaFe$_2$Se$_3$ ($x = 0$) and CsFe$_2$Se$_3$ ($x = 1$) show no metallization or superconductivity up to 30.0 and 17.0 GPa, respectively.
In contrast, BaFe$_2$S$_3$ shows metallization and superconductivity at 11 GPa\cite{takahashi2015pressure}.
This tendency is consistent with the report that photoemission spectroscopy at ambient pressure revealed a sizable energy gap for all these compounds, with the largest (smallest) energy gap in CsFe$_2$Se$_3$ (BaFe$_2$S$_3$) \cite{PhysRevB.91.014505}.
The Mott gap of BaFe$_2$S$_3$ is small enough to be suppressed by a pressure of 11 GPa, hence metallization is achieved.
On the other hand, the gaps of BaFe$_2$Se$_3$ and CsFe$_2$Se$_3$ are too large to be suppressed by the pressure of 30.0 and 17.0 GPa, respectively.

Next, we discuss the resistivity of the intermediate compounds.
The intermediate compounds, Ba$_{1-x}$Cs$_x$Fe$_2$Se$_3$, correspond to career-doped Mott insulator, so they could be metallic at ambient pressure.
However, our results show that they are insulators even at ambient pressure and low pressures below 10 GPa.
Additionally, the $x$ = 0.25 and 0.65 samples show metallic behavior at higher pressures, in contrast to the robust insulating behavior in the parent compounds.
These quantitatively different behaviors of resistivity at high pressure for the intermediate compounds from the parent compounds suggest intrinsically different mechanisms behind the insulating state.

To consider the origin of the insulating behavior in the intermediate compounds, their unique temperature dependence of resistivity can be helpful.
The resistivity can be well fitted by a one-dimensional variable-range-hopping (VRH) \cite{MOTT19681} type temperature dependence at ambient pressure \cite{PhysRevB.91.184416} and a two-dimensional weakly localized system-type dependence at high pressures.
Both temperature dependences are based on the same idea: Anderson localization \cite{PhysRev.109.1492}.
Therefore, based on the Anderson localization mechanism, we here offer the most plausible scenario for the behavior of resistivity in the intermediate compounds, Ba$_{1-x}$Cs$_x$Fe$_2$Se$_3$, as follows.
At ambient pressure, electrons are strongly localized because of the randomness of the potential, and the sample shows one-dimensional VRH-type resistivity at low temperatures.
The Arrhenius-type resistivity, instead of the VRH-type, is observed at room temperature (not shown), but this behavior does not contradict the Anderson localization scenario;  electrons can hop to much higher energy states in the nearest sites at high temperature \cite{NFMott}.
At high pressures, owing to the decrease in atom distances, hopping integral and dimensionality increase, and the random potential is effectively weakened and samples show metallic behavior.
However, the effective random potential is still valid as a perturbation, hence the conductivity of the $x$ = 0.25 and 0.65 samples shows a $\log T$ dependence at low temperatures, reflecting the weakly localized two-dimensional nature.
The complete suppression of the effect of the random potential upon the application of further higher pressure realizes the three-dimensional Fermi liquid behavior of the $x$ = 0.65 sample.
In addition, the finite $\gamma$ term of the specific heat for the $x$ = 0.25 and 0.65 samples \cite{PhysRevB.91.184416} supports the idea of Anderson localization; the $\gamma$ term could result from the density of states of electrons at the Fermi level.
The random potential could originate from a large difference of iconic radii between Ba (1.49 \AA ) and Cs (1.81 \AA) for Ba$_{1-x}$Cs$_x$Fe$_2$Se$_3$.
The low dimensionality originating from the two-leg ladder structure of Fe lattice enhances effective random potential.

The above Anderson localization scenario seems likely, but is a speculation.
To confirm this scenario, electrical resistivity and Hall effect measurements under a magnetic field should be important \cite{PhysRevB.22.5142}.
Photoemission spectroscopy is also meaningful for observing the density of states of electrons at the Fermi level.
Such tasks are out of the scope of the present study, and a future study along these lines is highly desired.

\section{Conclusion}
We have performed electrical resistivity measurements in the iron-based ladder compounds Ba$_{1-x}$Cs$_x$Fe$_2$Se$_3$ under high pressure using a cubic anvil press and a diamond anvil cell.
We achieved metallization in the $x$ = 0.25 and 0.65 samples at 11.3 and 14.4 GPa, respectively.
The samples show insulating behavior at low temperatures, indicating that the samples are weakly localized two-dimensional systems.
Fully metallic behavior was observed for the $x$ = 0.65 samples in the measured temperature range at 23.8 GPa, and the metallic state shows a three-dimensional Fermi liquid-like temperature dependence below 50 K.
We speculate that iron-based ladder compounds have two origins of the insulating state: the first is Mott gap for the parent compounds; the second is the random potential for intermediate compounds.

\section*{Acknowledgments}
This research was partly supported by a Grants-in-Aid for Scientific Research (Nos. 15F15023, 15H03681, 16H04019, 16H04007, and 23244068) from MEXT of Japan, the Nano-Macro Materials, Devices and System Research Alliance, and the Mitsubishi Foundation.

\bibliography{ref.bib}

\end{document}